\documentclass[aps, prl, letterpaper, amsmath, amssymb, notitlepage,showpacs,
superscriptaddress, twocolumn]{revtex4}

\usepackage{graphicx}

\newcommand{\eqb}{\begin{equation}}
\newcommand{\eqe}{\end{equation}}

\begin{document}

\title{Apex Exponents for Polymer--Probe Interactions}

\author{Michael Slutsky}
\affiliation{Department of Physics, Massachusetts Institute of
Technology, 77 Massachusetts Avenue, Cambridge, MA 02139, USA}

\author{Roya Zandi}
\affiliation{Department of Physics, Massachusetts Institute of
Technology, 77 Massachusetts Avenue, Cambridge, MA 02139, USA}
\affiliation{Department of Chemistry and Biochemistry, UCLA, Box 951569, Los Angeles, CA 90095-1569}

\author{Yacov Kantor}
\affiliation{School of Physics and
Astronomy, Tel Aviv University, Tel Aviv 69978, Israel}

\author{Mehran Kardar}
\affiliation{Department of Physics, Massachusetts Institute of
Technology, 77 Massachusetts Avenue, Cambridge, MA 02139, USA}

\date{\today}

\begin{abstract}
We consider self-avoiding polymers attached to the tip of an impenetrable probe.
The scaling exponents $\gamma_1$ and $\gamma_2$, characterizing the number of
configurations for the attachment of the polymer by one  end, or at its midpoint, 
vary continuously with the tip's angle.
These  apex exponents are calculated analytically by $\epsilon$-expansion,
and numerically by simulations in three dimensions.
We find that when the polymer can move through the attachment point, it typically
slides to one end;
the apex exponents quantify the entropic barrier 
to threading the eye of the probe.

\end{abstract}
\bibliographystyle{apsrev}
 \maketitle

There has been remarkable progress in recent years in nanoprobing
and single--molecule techniques.  These developments have had a
direct impact on biopolymer research producing a wealth of
beautiful results on DNA dynamics~\cite{Elbaum}, molecular
motors~\cite{Bustamante}, and protein/RNA
folding~\cite{Fernandez,Liphardt}.  Today it is possible to
measure statistical properties of a single macromolecule rather
than deducing them from experiments with solutions of many
polymers. This naturally leads to questions regarding the
theoretical limitations of these techniques, such as the effects
of microscopic probes on the measured properties of the
polymer. Consider, for instance, a polymer attached to the apex
of a cone--shaped probe (e.g. a micropipette or the tip of an
atomic force microscope~\cite{gaub,force}). What is the
configurational entropy for this system? Suppose that this probe
is a microscopic needle with a hole at the end. How hard is it to
thread a polymer through the needle's eye?

Quite generally, the number of configurations ${\mathfrak N}$ of
a polymer of length $N$ or, equivalently, of an $N$--step
self--avoiding walk (SAW), behaves as~\cite{degennesSC}
\begin{equation}\label{gammadef}
{\mathfrak N}= {\rm const}\times z^N  N^{\gamma-1}.
\end{equation}
The ``effective coordination number'' $z$, depends on microscopic
details, while the exponent $\gamma$ is `universal.'  Actually,
$\gamma$ does depend on geometric constraints which influence the
polymer at all length scales.  In particular, there are a number
of results demonstrating the variations of $\gamma$ for polymers
confined by wedges in two and three
dimensions~\cite{cardy_red,guttmann,debell,gsurface}: A SAW
anchored at the origin and confined to a solid wedge (in 3D) or a
planar wedge (in 2D) has an angle--dependent exponent $\gamma$
that diverges as the wedge angle 
vanishes.  A limiting
case which has been extensively studied, both
analytically~\cite{kosmas,douglas_kosmas} and
numerically~\cite{gsurface,debell}, is a SAW anchored to an
impenetrable surface, for which $\gamma \equiv \gamma_s =
0.70\pm0.02$~\cite{gsurface}.

\begin{figure}
\includegraphics[width=8cm]{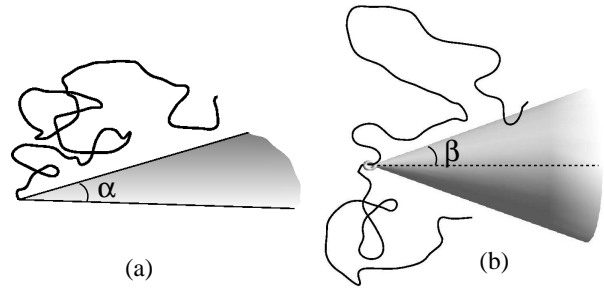}
\caption{\label{describe_defs} 
Configurations of a polymer near an obstacle: (a) attached to
the apex of a planar sector of angle $\alpha$; (b) threaded through the
eye of a cone with apex semi--angle $\beta$.}\label{fig:segment}
\end{figure}

To model the polymer--probe system, we consider a SAW attached to
the apex (tip) of an impenetrable obstacle (needle).  To avoid
introduction of an external length scale, we focus on obstacles
of scale--invariant shape, such as a planar slice (sector) of
angle $\alpha$ (Fig.~\ref{describe_defs}a), or a conical needle
of apex semi-angle $\beta$ (Fig.~\ref{describe_defs}b).  While
both geometries are natural extensions of the 2D wedge, they are
clearly different in three dimensions (and also distinct from the
3D wedge, which consists of two planes intersecting at a line).
The former excludes the polymer from the volume of a cone, while
the latter prevents it from crossing the surface of a slice.
Nonetheless, the resulting phenomenology is rather similar.
Indeed, one of the technical innovations of this paper is the
demonstration that many such geometries can be treated in the
same manner by an $\epsilon=4-d$ expansion focusing on the
interaction with a 2D surface.  The $\epsilon$--expansion, as
well as numerical simulations in 3D, shows that the exponent
$\gamma\equiv\gamma_1$ varies continuously with the apex opening
angles in Fig.~\ref{describe_defs}.  Continuously varying
exponents are rather uncommon in critical phenomena. In the
present case they arise from the interaction of two self-similar
entities, the polymer and the probe.

Another variant of this problem occurs when a polymer is attached
to the apex at its {\em midpoint}. This case is described by
Eq.~(\ref{gammadef}) with exponent
$\gamma\equiv\gamma_2$. More generally, let us denote by
${\mathfrak N}_2(N, N_1)$, the number of accessible configurations
for a polymer attached to the apex at an arbitrary monomer,
dividing it in two segments of lengths $N_1$ and $N_2 = N -
N_1$. If we allow the two segments to exchange monomers with each
other (which can be done by replacing a rigid attachment with a
slip--ring as depicted in the Fig.~\ref{describe_defs}b), then
the equilibrium configurations will be distributed with a weight
proportional to ${\mathfrak N}_2(N, N_1)$.  A natural
interpolation formula as a function of $\alpha$ (or $\beta$),
supported by the $\epsilon$-expansion at first order, is
\begin{equation}\label{ansatz}
{\mathfrak N}_2(N, N_1) \propto N^{c(\alpha)}[N_1(N-N_1)]^{c_1(\alpha)}.
\end{equation}
To get a feeling for this scaling relation, let us look at some
limits: When the probe is absent, we recover Eq.~(\ref{gammadef})
and $c(0) = \gamma_0 - 1$, where $\gamma_0\simeq 1.158$
describes the geometrically unconstrained SAW.  If the obstacle
is present but the two segments do not interact with each other,
then $c = 0$ and $c_1= \gamma_1 -1$. 
By fitting to the limits of $N_1\to0$ and $N_1\sim N_2$, 
we find $c_1= \gamma_2 - \gamma_1$ and
$c = 2\gamma_1 - \gamma_2 - 1$.  
Below, we estimate the exponents in Eq.(\ref{ansatz}) both analytically and numerically.  
For now, assuming Eq.~(\ref{ansatz}) holds, we see that if
$c_1< 0$, the maximum number of configurations is
realized when either $N_1$ or $N_2$ equals $N$. This brings us to
one of our main findings: No matter how small the apex angle,
we find  $c_1< 0$, i.e. the
most likely states have $N_1 \simeq N$ or $N_2 \simeq N$, with an
{\em entropy barrier} separating the two.  Threading a needle is
hard!

To treat the problem analytically, we start with the Edwards
\cite{edwards} model of a self-avoiding polymer, and add an
interaction with the obstacle.  In this formulation,
configurations of the polymer are described by ${\bf
r}(\tau)\in\Re^d$, where $\tau$ measures the position along the
chain, and are weighted according to the energy~\footnote{To simplify
notation, the monomer size is absorbed into a
redefinition of $N$, giving it  dimensions of [length]$^2$.}
\begin{eqnarray}\label{eq:perturb_ham}
{\cal H} &=& \frac{1}{2}\int_0^N \dot{\bf r}^2 ~d\tau +
\frac{v_0}{2}\int_0^N d\tau\int_0^N d\tau'\delta[{\bf r}(\tau) -
{\bf r}(\tau')] \nonumber\\
&+& g_0\int_{\mathcal{M}}d^2{\bf R}\int_0^N
d\tau\delta[{\bf r}(\tau) - {\bf R}].
\end{eqnarray}
The self-avoiding interaction is replaced by a `soft' repulsion of strength $v_0$.
In the same spirit, the impenetrable obstacle is replaced with a soft repulsion
of magnitude $g_0$.
The key observation is that in 3D the polymer can only sense the exterior of an
impenetrable obstacle, and will not care if its interior is hollow.
In generalizing to $d$-dimensions, we keep the dimensions of the
now softened exterior manifold (indicated by ${\bf R}\in{\mathcal M} $) as two.
The advantage of this choice is that both $g_0$ and $v_0$ have
the same bare dimensions, and in a perturbative scheme simultaneously 
become relevant in $d\leq 4$. 
We then analyze the model using a renormalization 
group (RG) scheme~\cite{kosmas,douglas_kosmas} which is a
modification of the conformation space RG \cite{oono3,freed_book}. 
The scaling exponents are calculated  using dimensional regularization in 
$d = 4-\epsilon$ dimensions to order $O(\epsilon)$.


\begin{figure}[htb]
\includegraphics[width = 8cm]{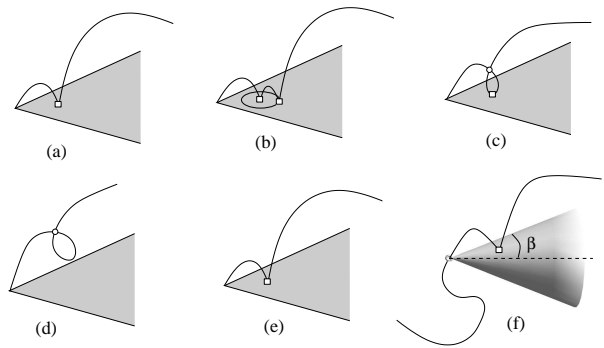}
\caption{\label{fig:renorm} Diagrams contributing to 
renormalization of $g$ to second order (a--c); to ${\mathcal
Z}$ in first order (d,e) at the apex of a slice; and
to ${\mathcal Z}_2$ in first order (f) at the eye of a conic needle.}
\end{figure}


It is customary to define non-dimensionalized (bare) coupling
constants $\tilde{v}_0 = v_0L^{\epsilon},~\tilde{g}_0 =
g_0L^{\epsilon}$ at a length scale $L$.  We also define the {\em
renormalized} coupling constants $v = Z^{-1}_v\tilde{v}_0$ and $g
= Z^{-1}_g\tilde{g}_0$, where the renormalization constants $Z_v$,
$Z_g$ are calculated perturbatively as series in $v$ and $g$.
Diagrams contributing to the renormalization of $g$ are shown in
Figs.~\ref{fig:renorm}a--c, and involve both interactions of the
polymer with the obstacle and with itself. The leading
singularity in $1/\epsilon$ comes from short distances and
therefore the leading correction to $g$ does not depend on the
overall shape of the obstacle.  This is true as long as it is
possible to draw a finite circle around points away from the apex
(see Fig.~\ref{fig:renorm}b), and is also why the interaction
with the obstacle becomes irrelevant in the degenerate limits of
$\alpha =\beta= 0$.  The self--interaction is uninfluenced by the
obstacle, and the renormalization constant $Z_v$ is the same as in the
unattached polymer~\cite{freed_book}. To first order in
$\epsilon$, the nontrivial fixed point of the RG flow is found to
be $(v^*, g^*) = (\pi^2\epsilon/2,3\pi\epsilon/4)$.
The fixed point thus depends only on the dimension of the constraining manifold,
but not on its shape~\cite{freed3,duplantier1}. However, the number of
accessible configurations does depend on the exact geometry as described below.

Consider first-order corrections to the partition function
${\mathcal Z}$ coming from the self--interaction
(Fig.~\ref{fig:renorm}d) and the interaction with the slice
(Fig.~\ref{fig:renorm}e).  Combining them and adding relevant
counterterms to eliminate poles in $\epsilon$, we obtain 
\eqb\label{Zslice}
{\mathcal Z} = 1 + \frac{1}{4\pi^2}(v^* - \alpha
g^*)\ln\left(\frac{2\pi N}{L^2}\right)
+ O(\epsilon^2).
\eqe
Comparing this with $N^{\gamma_1(\alpha)-1} = 1 +
(\gamma_1(\alpha)-1)\ln N +\cdots$, and substituting the fixed point
values $(v^*, g^*)$, we find
\eqb\label{alpha1rg}
\gamma_1(\alpha) = 1 +\frac{\epsilon}{8}\left(1 -
\frac{3\alpha}{2\pi}\right) + O(\epsilon^2).
\eqe
The above treatment is easily generalized to  a
polymer attached by its midpoint. For the number of
configurations, we observe that the contribution from the
interaction with the obstacle is doubled. Interaction between the
two halves of the polymer, however, makes no separate correction
and is already included.  
(Note that if we ignore the
obstacle and consider self-interactions only, we get a
``degenerate'' star polymer with two branches that is equivalent
to a linear polymer.)
Thus, for the calculation of $\gamma_2$
at order of $\epsilon$, we can add the separate contributions
from self-avoidance and avoidance of the obstacle; cross-terms
can only occur at higher orders.  This enables us to identify the scaling
exponent
\eqb\label{alpha2rg}
\gamma_2(\alpha) = 1 +\frac{\epsilon}{8}\left(1 -
\frac{3\alpha}{\pi}\right) + O(\epsilon^2).
\eqe
Repeating this argument for the slip--ring geometry, we find
\eqb
{\mathfrak N}_2 \propto N^{\epsilon/8}[N_1(N-N_1)]^{-3\alpha\epsilon/(16\pi)},
\eqe 
which confirms the Ansatz in Eq.~(\ref{ansatz}) to first order in
$\epsilon$. Note that in 3D we must have
$\gamma_2(2\pi)=2\gamma_s-1$. This equality does not hold in the
$\epsilon$--expansion. The reason is that in 3D, a complete plane
prevents two polymers on its opposite sides from interacting with
each other, whereas in 4D it does not.

It is straightforward to extend the above formalism to obstacles
of different shapes, such as the conical manifold with apex angle
$\beta$ (Fig.~\ref{describe_defs}b).  In counting the number of
configurations we obtain a result similar to Eq.~(\ref{Zslice}),
with $\alpha$ replaced by $2\pi\sin\beta$.  Since the fixed point
location is the same as before, this substitution in
Eqs.~(\ref{alpha1rg}-\ref{alpha2rg}) gives
\begin{eqnarray}\label{cone12}
\gamma_1^{\rm cone}(\beta) &=&  1 +\frac{\epsilon}{8}(1 -3\sin\beta) + O(\epsilon^2),\\
{\mathfrak N}_2^{\rm cone} (\beta) &\propto&
N^{\epsilon/8}[N_1(N-N_1)]^{-(3/8)\epsilon \sin\beta}.
\end{eqnarray}
The difference between the two geometries is thus
merely quantitative.  As another point of caution, we note that 
within our formalism, the polymer is free to occupy either
side of the hollow cone-- the partition sum is dominated by the
arrangement with the largest number of configurations. 
Thus the result for $\gamma_1(\beta)$ is valid only for $\beta\leq\pi/2$,
with a possibly more severe restriction for $\gamma_2$.

\begin{figure}
\includegraphics[width=7cm]{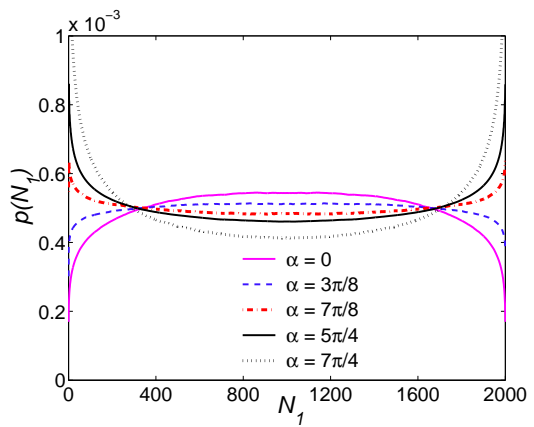}
\caption{\label{showeffect} The probability distributions $p(N_1)$ for 
 two non-interacting segments of lengths $N_1$ and $N-N_1$
 attached to the apex of a planar slice for diffrent values of angle $\alpha$. 
 The curves are the result of $10^9$ MC steps for $N=2000$.}
\end{figure}

The earlier discussion of `threading a needle' illustrates the essence of the 
method of ``entropic competition''~\cite{zandi,marcone}
which we employ to numerically 
estimate the exponents $\gamma_i(\alpha)$ in 3D.
The underlying idea  is
to sample the ensemble of different configurations of {\em two polymer
segments} which can exchange monomers and thus ``compete
entropically'' (e.g. in a slip--ring geometry). To calculate
$\gamma_1(\alpha)$, we prevent the two segments from
interacting with each other. The number of configurations 
for given $N$ and $N_2 = N- N_1$ is then
\begin{equation}\label{relative_conf}
{\mathfrak N}_1 \propto  [N_1(N - N_1)]^{\gamma_1(\alpha)-1},
\end{equation}
so that the resulting histogram for $N_1$ allows us to calculate
the exponent $\gamma_1(\alpha)$. In our simulations we consider
sectors with angles $\alpha=k\pi/8$, where $k=0,1,2,\dots,16$.
We work on a cubic lattice, with the excluded sector in the $x-y$
plane. Possible Monte Carlo (MC) moves include attempts to remove one
monomer from the {\it free} end of a randomly chosen polymer
segment and adding it to the {\it free} end of the other
segment; both segments also undergo random configuration changes
via pivoting~\cite{madrassokal}. 
Figure~\ref{showeffect} illustrates the dramatic effect of the
angle $\alpha$ on $p(N_1)$, the probability distribution function
(PDF) for the segment length $N_1$.  For small $\alpha$, the distribution
is peaked at the center while for $\alpha$ bigger than a critical
value $\alpha_c$, the maximum of the PDF moves to the sides. 
The numerical data from entropic competition suggest
$\alpha_c\approx 5\pi/8$, which is not too far from the first order
$\epsilon$--expansion result of
$\alpha_c = 2\pi/3$ in Eq.~(\ref{alpha1rg}). 

\begin{figure}
\includegraphics[width=8cm]{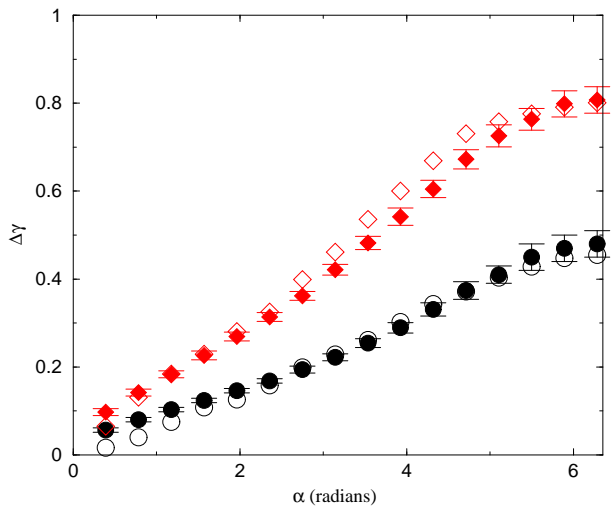}
\caption{\label{delta_gamma} Extrapolated values of the exponents
$\Delta\gamma_1=\gamma_0 - \gamma_1$ (circles) and 
$\Delta\gamma_2=\gamma_0 - \gamma_2$
(diamonds) as a function of sector angle $\alpha$ from `entropic
competition' (open symbols), and dimerization (full symbols). 
Error bars represent statistical uncertainties of
individual estimates of the exponents, as well as the
uncertainty in the extrapolation $N\to\infty$. }
\end{figure}

For the purpose of calculating $\gamma_2(\alpha)$, it
is necessary to include interactions between the segments.
Figure~\ref{delta_gamma} depicts variations of the exponents 
$\gamma_{1,2}(\alpha)$ 
obtained by fitting histograms from entropic competition,
such as in Fig.~\ref{showeffect},
to power-laws as in Eqs.~(\ref{relative_conf}) and (\ref{ansatz}).

It is instructive to compare the results
of entropic competition with those of a more established
procedure, such as
dimerization~\cite{dimerization1,dimerization2}.  The latter is a
quite efficient method~\cite{madrassokal}, in which an $N$-step
SAW is created by generating two $(N/2)$-step SAWs and attempting
to concatenate them.  We generated SAWs for $N=16,~32,\cdots
~,2048$, and by attempting to attach them to the end point of an
appropriate sector, measured a success probability $p_N$.  Let us
indicate the number of SAWs not attached to the sector by $A_0z^N
N^{\gamma_0-1}$, and those attached to the sector either (1) by
their ends, or (2) by their mid-point as $A_i z^N
N^{\gamma_i(\alpha)-1}$ ($i=1,2$ corresponds to the notation
introduced earlier).  Then, the ratio between the number of
configurations,
$p_N\equiv(A_i/A_0)N^{\gamma_i(\alpha)-\gamma_0}$, represents the
probability to attach an $N$--step polymer to a sector with angle
$\alpha$. Fitting a power law to this ratio thus 
provides a means of estimating  the exponent difference
$\Delta\gamma_i\equiv\gamma_0-\gamma_i$.

Using the dimerization method we generated $M=10^6$ SAWs;
for each   we checked whether there is an intersection 
with sectors at all tested  values of $\alpha$~\footnote{Since the
same ensemble of SAWs is used for all sector angles, the results
for various values of $\alpha$ are correlated.}. 
The error in the
measurement of $p_N$ is $\sqrt{p_N(1-p_N)/M}$. Since $p_N$
decreases with increasing angle of the sector, the relative error
in $p_N$, and an error in the estimates of $\Delta\gamma_i$,
increases with increasing $\alpha$. Nevertheless, 
we were able to obtain reasonable estimates of the exponent for
all values of $\alpha$, as shown in Fig.~\ref{delta_gamma} (full
symbols).

The two numerical approaches are in  very good agreement; 
error bars for `entropic competition' results being even smaller.  
For $\alpha=0$, our results deviate from zero beyond the statistical error
range. We believe this deviation to be a finite size effect, due to
discreteness of the lattice. As a check, we 
estimated $\Delta\gamma_{1,2}$  when the obstacle 
consists of the positive $x$-axis.  While asymptotically such a
situation corresponds to $\alpha=0$, and should lead to
$\Delta\gamma_i=0$, we obtained $\Delta\gamma_1=0.02$ and
$\Delta\gamma_2=0.05$. For $\alpha=2\pi$, we expect to have
$\Delta\gamma_1=\gamma_0-\gamma_s\approx 0.46$, and
$\Delta\gamma_2=\gamma_0-2\gamma_s+1\approx 0.76$;
our results are quite close to these estimates.

In summary, we consider configurations of a polymer attached to
the apex of a self-similar probe (at least on the scale of the
polymer size).  The geometric constraints imposed by the
impenetrable probe lead to exponents 
continuously with the apex angle.  Two such exponents are
associated with attachment of the polymer by one end or by a
mid-point.  Together, they determine if a mobile attachment point
is likely to be in the middle or slide to one side.  These apex
exponents are obtained analytically by an $\epsilon=4-d$
expansion, and through independent numerical schemes in $d=3$.
The $\epsilon$-expansion takes advantage of the marginality of
interactions of a polymer with a two dimensional manifold in four
dimensions, and can be applied to a variety of shapes.  The
numerical method of `entropic competition' is shown to be a
powerful tool in this context, comparable to or better than the
more standard dimerization approach.  The numerical and
analytical results are in agreement, and indicate the presence of
an entropic barrier that favors attachment of the polymer to the
apex at its end.  It would be interesting to see if these
predictions can be probed by single molecule experiments.

This work was supported by Israel Science Foundation grant
No. 38/02, and by the National Science Foundation grant
DMR-01-18213.
\bibliography{apex}

\end{document}